\documentclass{svproc}
\usepackage{url}

\usepackage{graphicx}
\usepackage{multirow}
\usepackage{biblatex}
\begin{document}
\mainmatter 

\title{Voice Activity Detection for Ultrasound-based Silent Speech Interfaces using Convolutional Neural Networks}

\titlerunning{Speech/Silence Detection for SSI} 

\author{Amin Honarmandi Shandiz \and László Tóth }
%
\authorrunning{Amin Shandiz and L. Tóth.} %

\tocauthor{A. Shandiz and L. Tóth}
%
\institute{University of Szeged, Institute of Informatics\\
\email{\{tothl, shandiz\}@inf.u-szeged.hu}
}

\maketitle  

\begin{abstract}

Voice Activity Detection (VAD) is not easy task when the input audio signal is noisy, and it is even more complicated when the input is not even an audio recording. This is the case with Silent Speech Interfaces (SSI) where we record the movement of the articulatory organs during speech, and we aim to reconstruct the speech signal from this recording. Our SSI system synthesizes speech from ultrasonic videos of the tongue movement, and the quality of the resulting speech signals are evaluated by metrics such as the mean squared error loss function of the underlying neural network and the Mel-Cepstral Distortion (MCD) of the reconstructed speech compared to the original. Here, we first demonstrate that the amount of silence in the training data can have an influence both on the MCD evaluation metric and on the performance of the neural network model. Then, we train a convolutional neural network classifier to separate silent and speech-containing ultrasound tongue images, using a conventional VAD algorithm to create the training labels from the corresponding speech signal. In the experiments our ultrasound-based speech/silence separator achieved a classification accuracy of about 85\% and an AUC score around 0.86.


\keywords{Silent Speech Interface, speech/silence classification, Voice Activity Detection, convolutional neural network}
\end{abstract}

\section{Introduction}

Voice Activity Detection (VAD) is an important component in many speech processing applications, for example automatic speech recognition (ASR)~\cite{atal1976pattern,lokhande2012voice} and speech enhancement~\cite{verteletskaya2010voice}. Its main role is to detect the presence or absence of speech~\cite{verteletskaya2010voice}, but sometimes it also involves a voiced/unvoiced decision~\cite{nirmalkar2016voiced}. Its application can not only significantly reduce the computational costs, but it may also influence the speech recognition accuracy~\cite{benyassine1997silence}.
In speech enhancement, VAD is used to identify frames which contain only noise to remove them from the signal~\cite{verteletskaya2010voice}. In machine learning-based speech synthesis,  removing noisy segments and pauses may help generate more accurate models.

In this paper we work with silent speech interfaces (SSI), which aim to convert articulatory signals to acoustic signals. In our case, the articulatory input corresponds to a sequences of ultrasound images that record the movement of the tongue during speaking. The goal is to convert this recording of the articulatory movement into  a speech signal. Many possible approaches exist for this, but the most recent studies all apply deep neural networks (DNNs) for this task~\cite{Tatulli2017,Csapo2020, Saha-ultra2speech,honarmandi2021improving,yu2021reconstructing}, and here we also apply neural structures that combine convolutional neural network (CNN) layers and recurrent layers such as the long short-term memory (LSTM) layer.

Similar to speech applications, voice activity detection may also be useful in SSI systems, for example for sparing with the energy consumption in wearable SSI devices. However, in this case creating VAD algorithms is much more difficult, as the lack of speech does not correspond to a lack of high-amplitude input signal. The tongue position is continuously being recorded and presented by the ultrasound imaging tool, even when the subject is not speaking.

In this paper we first demonstrate that the application of VAD may impact the accuracy of our SSI neural model, the speech synthesis network we apply, and even the evaluation metric we use. Then we implement a CNN to separate silence and speech frames based on the ultrasound tongue images, so we basically create a VAD algorithm that works with ultrasound images. Finally, we evaluate the performance of this algorithm experimentally, and we also compare the performance of our SSI framework with and without using the VAD algorithm.

The rest of the paper is organized as follows. In section~\ref{sec:framework}, we briefly present our SSI approach, the we talk about the problem of voice activity detection in~\ref{sec:VAD}. Then the experimental setup is described in section~\ref{sec:experimental} and the experiments are presented and discussed in Section~\ref{sec:res}. We close the paper with conclusions in Section~\ref{sec:conc}.

\section{The Ultrasound-Based SSI Framework}\label{sec:framework}

Our SSI system follows the structure recommended by Csapó et al.~\cite{Csapo2020} The input of the system is a sequence of ultrasound tongue images (UTI) that were recorded at a rate of 82 frames per second. The goal of the SSI system is to estimate the speech signal that belongs to the articulatory movement recorded in the ultrasound images, so the SSI system has to create a model for the articulatory-to-acoustic mapping. We estimate this mapping using deep neural networks (DNNs). For the training procedure, we assume that the speech signal was also recorded in parallel with the ultrasound video, as this speech signal serves as the training target. Also, to reduce the amount of training data required, we estimate a dense spectral representation instead of the speech signal itself. In practice it means that our SSI network converts the ultrasound video into a mel-spectrogram, and the output speech signal is generated from the mel-spectrogram using the WaveGlow neural vocoder~\cite{waveglow}. It was shown by Csapó et al. that this approach is feasible, and it can generate intelligible speech from a sequence of ultrasound images~\cite{Csapo2020}.

Here, we evaluate the accuracy of spectral regression by two simple metrics. One of them is the simply the Mean Squared Error (MSE) loss for the training of the neural network. The other one is the Mel-Cepstral Distortion (MCD) between the original speech signal and the speech signal reconstructed from the ultrasound input~\cite{MCD}, which is a popular metric of speech quality in speech synthesis~\cite{kominek}.

\section{Voice Activity Detection from Speech and from Ultrasound}\label{sec:VAD}

Its main role of Voice Activity Detection is to estimate the presence or absence of speech~\cite{verteletskaya2010voice}. In the simplest case, that is, in a quiet environment the lack of speech activity corresponds to silent parts in the input signal. Hence, the simplest VAD algorithms compare conventional acoustic features such as the signal's energy to a threshold~\cite{rabiner1978digital}. Exceeding the threshold signs the presence of speech (VAD=1), otherwise the signal is identified as silence (VAD=0). However, the task becomes much more difficult under noisy conditions.  The speech phones can be voiced and unvoiced, and the most difficult is to separate unvoiced parts from background noise. Thus many VAD algorithms extend the two-class classification to 3 classes, corresponding voiced, unvoiced and silent (V/UV/S) parts~\cite{nirmalkar2016voiced,verteletskaya2010voice}. Moreover, under noisy conditions simple acoustic features such as the signal's energy may be insufficient, so several more sophisticated features have been proposed. For example, the classic paper by Atal et al. performs the prediction based on five different measurements including zero-crossing rate, speech energy, correlation features, 12-pole linear predictive coding (LPC), and the energy of the prediction error~\cite{atal1976pattern}. Other authors used features such as the zero-crossing rate, spectral or cepstral features, empirical mode decomposition (EMD), and so on~\cite{atal1976pattern,haigh1993robust,nirmalkar2016voiced}. 
More recent studies apply pattern recognition and machine learning methods to perform the voiced/unvoiced decision~\cite{moattar2010new,qi1993voiced,qi2004novel,deng2007voiced}. Mondal et al. applied clustering over temporal and spectral parameters to implement their VAD~\cite{mondal2015clustering}.

\begin{figure}[ht]
\centering
\includegraphics[width=0.6\textwidth]{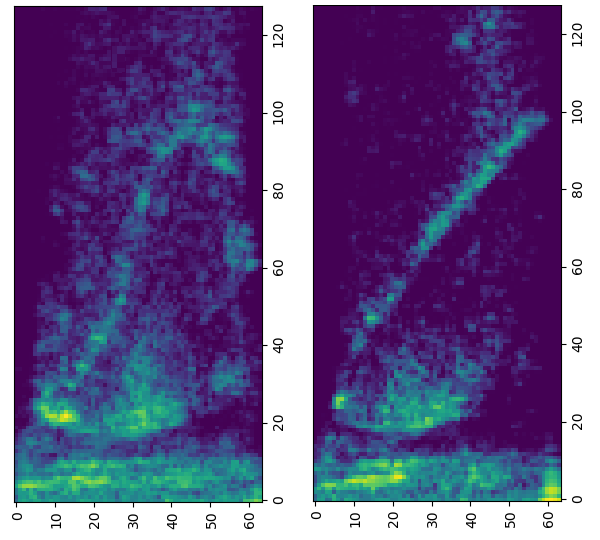}
\caption{Two UTI examples from the database, one for a speech frame (left) and one for a silent frame (right).} \label{fig_UTI}
\end{figure}

While there are a lot of studies on voice activity detection from speech, the input of our SSI system consists of ultrasound tongue images (UTIs). Fig~\ref{fig_UTI} shows two examples of the tongue position recorded by the device, when the subject is speaking and when he is not -- the diagonal light stripes in the images correspond to the tongue of the speaker. Clearly, separating ultrasound frames
that correspond to speech from those that represent silence seems to be more difficult than in the case of speech signals. In the following we train a CNN to perform the voiced/invoiced classification using such ultrasound images. As the structure of this VAD-CNN and the network that we apply for the SSI task are very similar, we describe them together in the next Section.

\section{Experimental Setup}\label{sec:experimental}

\subsection{The Ultrasound Dataset}\label{sec:data}

For the experiments we used English TAL corpus~\cite{ribeiro2020tal}. It contains parallel ultrasound, speech and lip video recordings from 81 native English speakers, and we used just the TaL1 subset which contains recordings from one male native speaker. 
We partitioned his files into training, testing and validation sets using 1015, 24 and 50 files,  respectively. To preprocess the ultrasound images we applied minmax normalization to the [-1,1] range, and resized the images to 64*128 pixels using bicubic interpolation. As regards the normalization of the speech mel-spectrogram features, we tried different normalization techniques, but we got the best results with the standard mean-deviance normalization (standardization). These 80 mel-spectrogram coefficients served as the training target values for the SSI network.

\subsection{CNNs for the SSI and for the VAD Task}

Convolutional Neural Networks are currently the most popular tool in image recognition, as they proved very powerful in extracting complex features from images by creating very deep network architectures~\cite{krizhevsky2012imagenet}. Standard CNNs convolve 2D filters with the images, but when the input is a video or 
a time series, CNNs can be extended to 3D by considering time as the third dimension~\cite{zhao2017pooling,ji20123d}. Recurrent neural networks such as the LSTM can also be very effective in extracting and combining temporal information from a sequential input~\cite{schmidhuber1997long}. However, these networks are known to be slow, so variants such as the quasi-recurrent neural network have been proposed~\cite{bradbury2016quasi}. This is why several authors apply 3D-CNNs to substitute recurrent layers when applying CNNs to a sequence of images~\cite{zhao2017pooling}. Here, we experiment with two neural network configurations in our SSI framework, that is, to estimate a speech mel-spectrogram from a sequence of ultrasound images. The first network is a 3D-CNN, following the proposal of Tóth et al.~\cite{toth20203d}.  The second configuration combines the 3D-CNN layers with and additional BiLSTM layer, as it may be more effective in aggregating the information along the time axis.
The structure of the two networks is compared in Table~\ref{tbl_cnn}. The input for both networks is the same, a short sequence of adjacent UTI frames. The output corresponds to the 80 mel-spectral coefficents that has to be estimated for the WaveGlow speech synthsis step, and the network is trained to minimize the MSE between the target and the output spectral vectors. 
\begin{table}[ht]
\centering
\caption{The structure of the 3D-CNN and the 3D-CNN + BiLSTM networks in Keras for the SSI task. The differences are shown in bold.}
\begin{tabular}{|l|l|}
\hline
\textbf{Conv3D}                                                                                                                   & \textbf{Conv3D+BiLSTM}                                                                                                              \\ \hline
\begin{tabular}[c]{@{}l@{}}Conv3D(30,(5,13,13),strides=(5,2,2))~~\\ Dropout(0.2)\end{tabular}                                              & \begin{tabular}[c]{@{}l@{}}Conv3D(30,(5,13,13),strides=(5,2,2))\\ Dropout(0.2)\end{tabular}                                  \\
\begin{tabular}[c]{@{}l@{}}Conv3D(60,(1,13,13),strides=(1,2,2))\\ Dropout(0.2)\\ MaxPooling3D(poolsize=(1,2,2))\end{tabular}             & \begin{tabular}[c]{@{}l@{}}Conv3D(60,(1,13,13),strides=(1,2,2))\\ Dropout(0.2)\\ MaxPooling3D(poolsize=(1,2,2))\end{tabular} \\
\begin{tabular}[c]{@{}l@{}}Conv3D(90,(1,13,13),strides=(1,2,1))\\ Dropout(0.2)\end{tabular}                                              & \begin{tabular}[c]{@{}l@{}}Conv3D(90,(1,13,13),strides=(1,2,1))\\ Dropout(0.2)\end{tabular}                                  \\
\begin{tabular}[c]{@{}l@{}}Conv3D(85,(1,13,13),strides=(1,2,2))\\ Dropout(0.2)\\ MaxPooling3D(poolsize=(1,2,2))\end{tabular}             & \begin{tabular}[c]{@{}l@{}}Conv3D(85,(1,13,13),strides=(1,2,2))\\ Dropout(0.2)\\ MaxPooling3D(poolsize=(1,2,2))\end{tabular} \\
\textbf{\begin{tabular}[c]{@{}l@{}}Flatten()\\ Dense(500)\\ Dropout(0.2) \end{tabular}}                                                  & \textbf{\begin{tabular}[c]{@{}l@{}}Reshape((5, 340))\\Bidirectional(LSTM(320,\\return\_sequences=False))\end{tabular}}                                       \\
Dense(80,activation='linear')                                                                                                            & Dense(80,activation='linear')                                                                                                \\ \hline\end{tabular}
\label{tbl_cnn}
\end{table}

\begin{table}[ht]
\centering
\caption{The structure of the 2D-CNN used for classification of speech/silent ultrasound images.}
\begin{tabular}{|l|}
\hline
\textbf{Conv2D}                                                                                                                  \\ \hline
\begin{tabular}[c]{@{}l@{}}Conv2D(32, (3, 3), padding='same', Activation='relu')\\ MaxPooling2D((2,2))\end{tabular}              \\
\begin{tabular}[c]{@{}l@{}}Conv2D(64, (3, 3), padding='same', activation='relu')\\ MaxPooling2D((2,2))\end{tabular}              \\
\begin{tabular}[c]{@{}l@{}}Conv2D(128, (3, 3), padding='same', Activation='relu')\\ MaxPooling2D((2,2))\\ Flatten()\end{tabular} \\
Dense(128, activation='relu'))                                                                                                   \\
Dense(1, activation='sigmoid')                                                                                                   \\ \hline
\end{tabular}
\label{tbl_classification}
\end{table}

We also trained a CNN to perform VAD from the ultrasound images. In this case we applied simple frame-by-frame training, so the input consisted of a single image, and we applied a 2D-CNN to classify the actual frame as silence or speech (Si/Sp). The architecture of this network is shown in Table~\ref{tbl_classification}. The network has just a single output that estimates the probability of the actual frame containing silence. This network was trained with the binary cross-entropy loss function.

\section{Results and Discussion}\label{sec:res}

\subsection{The Impact of VAD on the MCD Metric and on Speech Synthesis}

\begin{figure}[ht]
\centering
\includegraphics[width=1.0\textwidth]{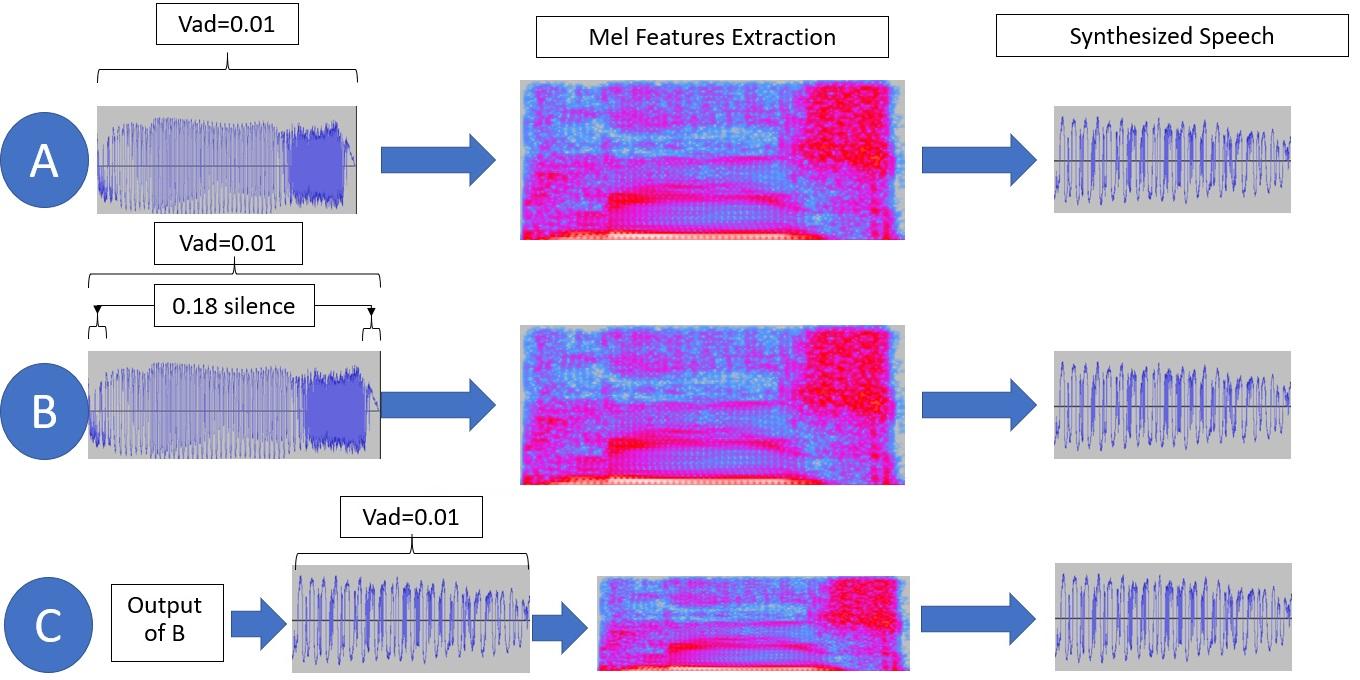}
\caption{Illustration of the experimental configurations applied in Table~\ref{vad-mcd}.} \label{figa}
\end{figure}

In the first experiment our goal was to demonstrate how the application of VAD may influence our results. Notice that this first experiment did not involve  SSI: we simply converted the speech signals to mel-spectrograms, and then reconstructed them using WaveGlow. We applied the Mel-Cepstral Distortion (MCD) metric to quantify the difference between the original and the reconstructed speech signals. For voice activiry detection and silence removal we used the VAD implementation available from WebRTC~\cite{WebRtc}. As shown in the first two rows of Table~\ref{vad-mcd}, retaining longer silent parts before and after the speech signal does influence MCD. We performed more experiments with preserving longer silent parts, and Fig~\ref{figa} shows that we obtained consistently increasing MCD values. This result indicates that MCD is sensitive to the amount of silence in the input. We should mention that some authors explicitly exclude the non-speech frames from the calculation of MCD~\cite{kominek}, but most papers do not clearly describe this step. 

In the third row of Table~\ref{vad-mcd} we show one more experiment where we performed two analysis-synthesis steps. Theoretically, the analysis and synthesis steps should be the perfect inverse of each other, so experiment c) should give the same result as experiment a). However, we obtained a slightly different MCD value. The probable explanation is that the WaveGlow speech synthesis network does not give a prefect reconstruction, and it is sensitive to certain parameters such as the duration of the silent parts or the positioning of the input windows.

\begin{table}
\caption{MCD values of the speech analysis-synthesis process when applying silence removal with three different VAD configurations.}
\begin{tabular}{|l|l|l|l|l|l|l|l|}
\hline
  \textbf{Configuration}   & \textbf{MCD}  \\ \hline
A: VAD (window length = 10 ms)  & 1.55                     \\ \hline
B: VAD (window length = 10 ms), plus keeping 180ms silence at both ends~~ & 2.03                      \\ \hline
C: applying A after B & 1.34                       \\ \hline
\end{tabular}
\label{vad-mcd}
\end{table}

\begin{figure}[ht]
\centering
\includegraphics[width=0.7\textwidth]{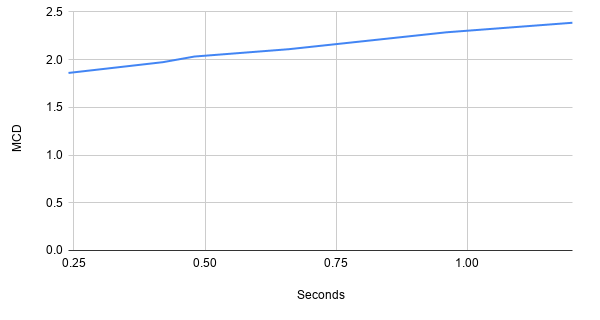}
\caption{Retaining more silence increases the MCD.} \label{figb}
\end{figure}

\subsection{The impact of VAD on the SSI}

In the second experiment we evaluated how the application of VAD on the training corpus influences the performance of our SSI system. In these UTI-to-speech conversion experiments we used the ultrasound data set presented in Section~\ref{sec:data}, and the two network configurations we described in Table~\ref{fig_CNN}.  Both models were trained using the Adam optimizer with a initial learning rate of 0.0002. We repeated the experiment with using the original training data and with removing most of the silent parts from the speech signals using the WebRTC VAD implementation. 

In Table~\ref{last-result} we report the MSE of the training process and the MCD values obtained from comparing the originals speech signals with those synthesized from the UTI input. The first thing we may notice is that the 3D-CNN+BiLSTM network produced much lower MSE rates and also slightly lower MCD errors. This shows the clear advantage of using a BiLSTM layer instead of a simple Dense layer. Second, the MCD scores are much higher in this case than in the previous experience. This is because there we worked with the original spectrograms, so the reported MCD values of 1.3-2.0 were caused by the inaccuracy of the WaveGlow neural vocoder. Here, the spectrograms were estimated from the UTI images, so the errors of our spectral estimation network and the WaveGlow network add up. Our best MCD score of 3.08 corresponds to a low-quality but intelligible speech~\cite{Csapo2020}. In comparison, Ribeiro et al. obtained an MCD score of 2.99 on the same corpus using more sophisticated encoder-decoder networks~\cite{ribeiro2020tal}.

As the last observation, we can see that retaining more silence in the corpus results in significantly lower MSE rates during training. The trivial explanation is that estimating silence is much easier for the network than estimating the spectrum of various speech sounds. However, the reduction of the MSE is misleading, as the MCD values on the test set have increased. Although this increase is not significant, it warns us that adding more training samples from a single class -- especially, from a trivial class -- may have a detrimental effect on the performance of a DNN, as it might shift the focus of training. 

\begin{table}[ht!]
\centering
\caption{Evaluation metrics of the SSI system after training the models with removing or retaining silence in the speech data.}
\begin{tabular}{|l|l|l|l|l|l|l|}
\hline
               & \multicolumn{3}{l|}{~~Removing silence by VAD} & \multicolumn{3}{l|}{~VAD + keeping 180ms silence} \\ \hline 
               & ~MSE (dev)~             & ~MSE (test)~             & ~MCD~             & ~MSE (dev)                      & ~MSE (test)                      & ~MCD~                       \\ \hline
Conv3d         & ~0.46                & ~0.45                & ~3.20            & ~0.30                        & ~0.33                         & ~3.29                      \\ \hline
Conv3d+BiLSTM & ~0.39                & ~0.42                 & ~3.08            & ~0.259                        & ~0.29                         & ~3.13                      \\ \hline
\end{tabular}
\label{last-result}
\end{table}

\subsection{Classification of Speech and Silence from Ultrasound}\label{sec_class}

\begin{figure}[ht]
\centering
\includegraphics[width=0.75\textwidth]{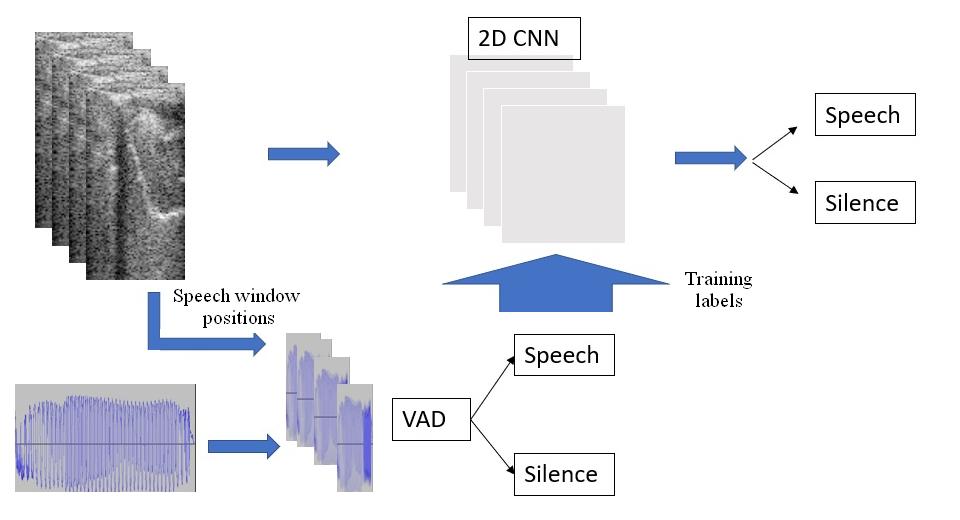}
\caption{Illustration of obtaining the VAD training labels and training the 2D-CNN for silence/speech classification.} \label{fig_CNN}
\end{figure}

In the previous experiment the VAD algorithm was executed with the speech signal. However, in practical SSI applications the speech signal might not be available, so we should be able to perform the voice activity detection from the ultrasound input. With this aim, we performed experiments to separate speech and silence frames of the ultrasound video using the 2D-CNN presented earlier in Table~\ref{tbl_classification}. The training labels for this 2-class classification process were obtained as follows (see also Fig~\ref{fig_CNN}).
As we have the synchronized speech signals for the ultrasound videos, we first identified the speech frames that belong to each ultrasound image based on the ultrasound frame rate. We split the speech signal into frames and fed it to the speech VAD function to decide about the speech/silence label of each image. We used these target labels with the ultrasound images as input to train the 2D-CNN for ultrasound-based voice activity detection. We used the ReLU activation function for all layers except the last layer which applies a sigmoid activation function to produce an output value between 0 and 1. For training we used SGD optimization with an initial learning rate of 0.001. We extracted the speech/silence training labels from the same train, development and test files as earlier, and the amount of speech labels was approximately 2-3 times more than the number of frames labelled as silence.

The evaluation metrics for this 2-class task are shown in Table~\ref{tbl_acc}. Besides the usual classification accuracy, we also report the precision and recall values which shows that the two classes were slightly imbalanced.
This is also reflected by the confusion matrices which can be seen in table~\ref{tbl_cm}. Thus, we also evaluated the AUC score based on the ROC of the classifier, which gave 0.89 for the development and 0.86 for the test set, respectively. Finally, we also mention that the F1 measure of 0.9 is also very good, and it could even be slightly improved by fine-tuning the decision threshold (which we did not adjust here). We also display the Cohen's Kappa values, which is a preferred metric in the case of imbalanced classes.

\begin{table}[ht]
\centering
\caption{}
\begin{tabular}{|l|l|l|}
\hline
             & \textbf{Dev set} & \textbf{Test set} \\ \hline
Accuracy     & 0.87             & 0.852             \\ \hline
recall       & 0.94              & 0.95             \\ \hline
precision    & 0.877             & 0.864            \\ \hline
F1           & 0.91               & 0.9             \\ \hline
ROC AUC      & 0.894             & 0.859            \\ \hline
Cohen's Kappa & 0.672              & 0.57            \\ \hline
\end{tabular}
\label{tbl_acc}
\end{table}
\begin{table}[ht]
\centering
\caption{Confusion Matrices for the silence/speech classification task for the development and test sets. }
\begin{tabular}{|c|l|l|l|l|l|l|l|}
\hline
\multicolumn{4}{|c|}{Dev Data}                          & \multicolumn{4}{c|}{Test Data}                            \\ \hline
\multicolumn{1}{|l|}{}  & \multicolumn{3}{c|}{Predicted} &                         & \multicolumn{3}{c|}{Predicted} \\ \hline
\multirow{3}{*}{Actual} &          & Negative & Positive & \multirow{3}{*}{Actual} &          & Negative & Positive \\ \cline{2-4} \cline{6-8} 
                        & Negative & 2850     & 1302     &                         & Negative & 1671     & 1268     \\ \cline{2-4} \cline{6-8} 
                        & Positive & 502      & 9295     &                         & Positive & 418      & 8096     \\ \hline
\end{tabular}
\label{tbl_cm}
\end{table}

\subsection{Replacing Speech-VAD by UTI-VAD}\label{sec_sputi}

\begin{table}[ht!]
\centering
\caption{Training the SSI system with removing or retaining silence from the data using the ultrasound-based VAD algorithm.}
\begin{tabular}{|l|l|l|l|l|l|l|}
\hline
              & \multicolumn{3}{l|}{~Removing silence by VAD~} & \multicolumn{3}{l|}{~VAD + keeping 180ms silence~} \\ \hline
              & ~MSE(dev)~     & ~MSE(test)~     & ~MCD~      & ~MSE(dev)~      & ~MSE(test)~      & ~MCD~       \\ \hline
conv3D        & ~~0.436        & ~~0.428         & ~~3.15     & ~~0.38          & ~~0.27           & ~~3.28      \\ \hline
Conv3D+BiLSTM & ~~0.393        & ~~0.41          & ~~3.05     & ~~0.35          & ~~0.26           & ~~3.12     \\ \hline
\end{tabular}
\label{utivad-result}
\end{table}

Finally, we repeated the experiment of Table~\ref{last-result}, but this time using the UTI-based VAD algorithm instead of the speech VAD. As the results in Table~\ref{utivad-result} show, in this case we obtained even slightly better MSE rates than earlier with the standard VAD function. The MCD values are basically equivalent with those obtained earlier, and the slight advantage of training the system with the removal of the long silent parts remained. In summary, we can say that our ultrasound-based VAD algorithm performed similarly to the standard, speech-based VAD algorithm in this experiment.

\section{Conclusion}\label{sec:conc}
Here we showed that -- similar to voice activity detection for speech -- ultrasound images can also predict and discriminate between Si/Sp segments. We estimated out training labels based on the parallel speech recording using a public VAD implementation. Our classifier attained a promising accuracy of 86\% in discriminating frames of silence and speech. We also showed that preserving too much silence in the training set can influence both the training of the model and the quality of the generated speech. In the future, we plan to apply our VAD technique as a method of silence removal as an initial step before feature extraction. That it, we window the speech signal in synchrony with the ultrasound frames and feed them to the VAD, and perform the subsequent feature extraction steps for synthesizing speech or other related tasks by using only the frames retained by VAD. 

\section{Acknowledgements}


\printbibliography

\end{document}